\documentclass{aa}  
\usepackage{graphicx}
\usepackage[varg]{txfonts}
\usepackage[svgnames]{xcolor}
\usepackage[colorlinks=true,linkcolor=blue,urlcolor=blue,citecolor=blue]{hyperref}

\usepackage{natbib}
\bibpunct{(}{)}{;}{a}{}{,}

\begin{document}

   \title{Exploring the connection between atmosphere models and evolution models of very massive stars}

   \author{J. Josiek\inst{\ref{inst:ari}}
          \and
          A.A.C. Sander\inst{\ref{inst:ari}}
          \and
          M. Bernini-Peron\inst{\ref{inst:ari}}
          \and
          S. Ekström\inst{\ref{inst:obsge}}
          \and
          G. Gonz\'alez-Tor\`a\inst{\ref{inst:ari}}
          \and
          R.R. Lefever \inst{\ref{inst:ari}}
          \and
          N. Moens \inst{\ref{inst:ivs}}
          \and 
          V. Ramachandran \inst{\ref{inst:ari}}
          \and
          E.C. Sch{\"o}sser \inst{\ref{inst:ari}}}

   \institute{Zentrum f{\"u}r Astronomie der Universit{\"a}t Heidelberg, Astronomisches Rechen-Institut, M{\"o}nchhofstr. 12-14, 69120 Heidelberg, \\Germany\label{inst:ari}
   \\ \email{joris.josiek@uni-heidelberg.de}
    \and
    Department of Astronomy, University of Geneva, Chemin Pegasi 51, 1290 Versoix, Switzerland\label{inst:obsge}
    \and
    Instituut voor Sterrenkunde, KU Leuven, Celestijnenlaan 200D, 3001 Leuven, Belgium \label{inst:ivs}             
    }

   \date{Received 10 February 2025 / Accepted 10 April 2025}

  \abstract{
  Very massive stars (VMS) are stars that are born with masses of more than 100\,$M_\odot$. Despite their rarity, their dominance in the integrated light of young stellar populations and their strong stellar feedback make them worthwhile objects of study. Their evolution is dominated by mass loss, rather than interior processes, which underlines the significance of an accurate understanding of their atmosphere. Yet, current evolution models are required to make certain assumptions on the atmospheric physics which are fundamentally incompatible with the nature of VMS.}
  {In this work, we aim to understand the physics of VMS atmospheres throughout their evolution by supplementing the structure models with detailed atmosphere models capable of capturing the physics of a radially-expanding medium outside of local thermodynamic equilibrium. An important aspect is the computation of atmosphere models reaching into deeper layers of the star, notably including the iron-opacity peak as an important source of radiative driving. From this we investigate the importance of the seemingly arbitrary choice of the lower boundary radius of the atmosphere model.}
  {We use the stellar evolution code GENEC to compute a grid of VMS models at solar metallicity for various masses. This grid implements a new prescription for the mass-loss rate of VMS. We then select the 150\,$M_\odot$ track and compute atmosphere models at 16 snapshots along its main sequence using the stellar atmosphere code PoWR. For each snapshot, we compute two atmosphere models connected to the underlying evolutionary track at different depths (below and above the hot iron bump), sourcing all relevant stellar parameters from the evolutionary track itself.}
  {We present two spectroscopic evolutionary sequences for the $150\,M_\odot$ track connected to an atmosphere model at different depths. Furthermore, we report on important aspects of the interior structure of the atmosphere from the perspective of atmosphere vs. evolution models. Finally, we present a generalized method for the correction of the effective temperature in evolution models and compare it with results from our atmosphere models.} 
  {The different choice of the connection between structure and atmosphere models has a severe influence on the predicted spectral appearance, which constitutes a previously unexplored source of uncertainty in quantitative spectroscopy. The simplified atmosphere treatment of current stellar structure codes likely leads to an overestimation of the spatial extension of very massive stars, caused by opacity-induced sub-surface inflation. This inflation does not occur in our deep atmosphere models, resulting in a discrepancy in predicted effective temperatures of up to $20$\,kK. Future improvements with turbulence and dynamically-consistent models may resolve these discrepancies.}

              
   \maketitle

\section{Introduction}

Massive stars ($M_\mathrm{ini}\gtrsim 8\,M_\odot$) are powerhouses in the Universe, driving the evolution of their host galaxies through extreme radiative, chemical, and mechanical feedback. Among them, very massive stars (VMS, $M_\mathrm{ini}\gtrsim 100\,M_\odot$) are the most extreme specimen, shining millions of times brighter than their low-mass counterparts and producing mass-loss rates on the order of $10^{-4}\,M_\odot\,/\mathrm{yr}$. Although initially predicted to be very rare by the initial mass function \citep[IMF,][]{Salpeter1955}, more recent evidence suggests that they are greater in number than expected \citep{Schneider2018, Ramachandran2018, Vink2018, Zeidler2017}. Moreover, their powerful impact on their surroundings means that, despite their rarity, an accurate understanding of their physics is required for an accurate interpretation of the role of massive star populations as a whole \citep[e.g.,][]{Crowther2016, Higgins2023, Upadhyaya2024}. 

In terms of evolution, VMS are generally considered a small subcategory at the upper end of the massive star regime. However, there are several indicators that set VMS apart from other massive stars, encouraging us to treat them as their own distinct class of star. Firstly, unlike other massive stars, most models suggest that VMS lose enough mass while still on the main sequence (MS) to transition directly into the hydrogen-poor Wolf-Rayet (WR) regime before core hydrogen exhaustion, bypassing the blue/red supergiant phase of evolution \citep[e.g.,][]{Ekstroem2012, Koehler2015, GormazMatamala2025}.
Spectroscopically, most observed VMS appear as WNh stars, morphologically extending from the early O-star regime \citep[e.g.,][]{Crowther2011,Bestenlehner2014}. VMS are also potential progenitors of pair-instability supernovae (PISNe) and the still elusive intermediate-mass black holes \citep{Gabrielli2024, Spera2017}.

From a modeling perspective, VMS still present significant challenges due to their complex physics. Stellar evolution models, which assume hydrostatic equilibrium and a simplified, optically thin atmosphere, are incompatible with the expanding atmospheres and strong winds expected and observed for VMS. Consequently, the effective temperatures predicted from structure models and the resulting evolution in the Hertzsprung-Russell diagram (HRD) are questionable. Additionally, the evolution of VMS is often modeled with mass-loss rates extrapolated from a lower-mass and lower-luminosity regime, introducing further uncertainties which then propagate through the entire evolution \citep{Graefener2021, Josiek2024}. 

In this work, we aim to characterize the structure and appearance of VMS through the combination of evolution models with detailed atmosphere models. This allows us to exploit the strengths of both modeling techniques and constitutes a first step towards understanding VMS in a more physically consistent way. Furthermore, we investigate the robustness of this approach by exploring different boundary conditions between the atmosphere model and the underlying structure models along the evolutionary path.

The paper is laid out as follows: Sect.\,\ref{sec:evol-models} introduces our evolution models and presents a new grid of VMS models with updated mass-loss rates. Sect.\,\ref{sec:atm-models} presents our atmosphere models, detailing the methodology we use to post-process the evolutionary tracks with the atmosphere code. We then discuss the resulting spectroscopic evolution of VMS in Sect.\,\ref{sec:shallow_models} and examine their atmospheric structure in Sect.\,\ref{sec:structure}. A summary of our findings is given in Sect.\,\ref{sec:summary}.

\section{Stellar evolution models}
\label{sec:evol-models}

\subsection{Physical ingredients}

In this work, we model VMS using the Geneva stellar evolution code \citep[GENEC,][]{Eggenberger2008, Ekstroem2012}. The basic input physics are equivalent to those used in the grid by \citet{Ekstroem2012}, notably using abundances from \citet{Asplund2005} \citep[except for Ne, which is from][]{Cunha2006}, and applying convection using a step-overshoot scheme with $\alpha=0.1$ with the Schwarzschild criterion.

We implement the recent dedicated VMS mass-loss scheme from \citet{Sabhahit2022}, which initially applies the standard mass-loss recipe for hot stars from \citet{Vink2001} and switches to an increased mass-loss rate as the star approaches the Eddington limit.

\subsection{Calculated models}

\begin{figure}
    \centering
    \includegraphics{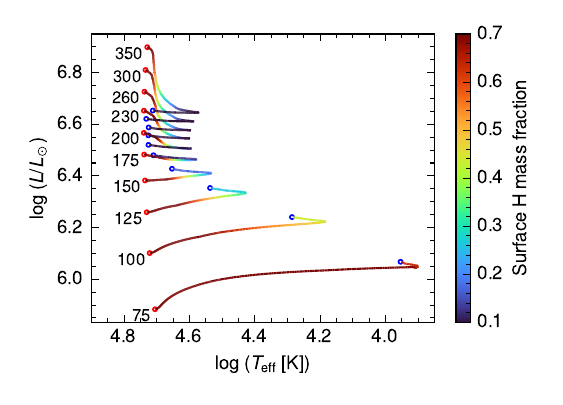}
    \caption{Hertzsprung-Russell diagram showing the evolution of the non-rotating models of potential VMS computed with GENEC using the updated mass-loss scheme. Each track is labeled with the corresponding initial mass (in $M_\odot$). The red and blue markers represent the ZAMS and TAMS, respectively. The color of the tracks indicates the surface hydrogen content across the evolution.}
    \label{fig:genec-full-hrd}
\end{figure}

Figure \ref{fig:genec-full-hrd} shows the grid of evolutionary tracks of our solar-metallicity, non-rotating VMS models computed with the updated mass-loss scheme in GENEC. All models were computed until the terminal-age main sequence (TAMS), which we define here as the moment when the central hydrogen mass fraction drops below $10^{-5}$.

\begin{figure}
    \centering
    \includegraphics{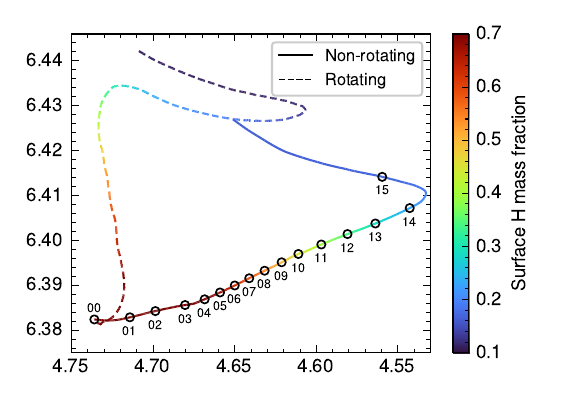}
    \caption{Hertzsprung-Russell diagram showing the evolution of the $150\,M_\odot$ models computed with GENEC using the new mass-loss scheme. The initial velocity of the rotating model was set at $10\%$ of the critical velocity. The markers (labeled 00--15) represent the evolution snapshots used to compute atmosphere models with PoWR along the non-rotating track. The color of the tracks indicates the evolving surface hydrogen content of the models.}
    \label{fig:genec-hrd}
\end{figure}

As a prototype of VMS evolution, we now focus specifically on the $150\,M_\odot$ model. According to typical evolutionary (i.e., non-spectroscopic) surface abundance criteria, this star is expected to evolve from O to WR type while still on the MS \citep[e.g.,][]{Yusof2013,Josiek2024}, allowing us to study the WR transition while avoiding uncertainties associated with post-MS evolution. Fig.\,\ref{fig:genec-hrd} shows the MS track of the non-rotating $150\,M_\odot$ model from our grid, as well as a model initially rotating at $0.1v_\mathrm{crit}$ computed with the same mass-loss prescription for comparison.

The rotating model inflates much less along the MS than the non-rotating one, despite having a larger core-to-envelope ratio. The inflation taking place for these models on the MS is mainly caused by an opacity effect close to the surface. The rotating model, having more efficient mixing, evolves almost chemically homogeneously with a significantly higher surface helium mass fraction than the non-rotating model. Because of this, the Thompson scattering opacity in the atmosphere is reduced and there is less inflation. 

\begin{figure}
    \centering
    \includegraphics{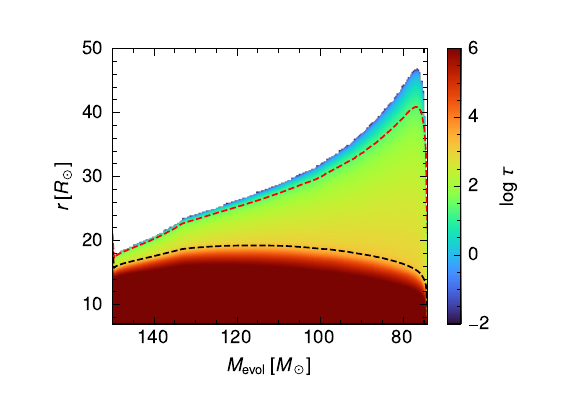}
    \includegraphics{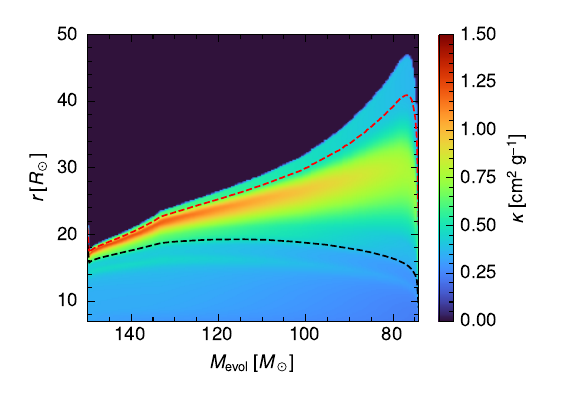}
    \caption{Evolution of the interior atmospheric structure of the continuum optical depth (top) and the total Rosseland mean opacity (bottom) of a non-rotating $150\,M_\odot$ star during the MS. The dashed red line shows the location of $\tau = 20$, and the dashed black line marks a local density of $10^{-7}\,\mathrm{g}\,\mathrm{cm}^{-3}$. These represent the connection boundary for the atmosphere models outlined in Sect.\,\ref{sec:powr-models}.}
    \label{fig:interior_tau_evol}
\end{figure}

Figure \ref{fig:interior_tau_evol} underlines that the entire apparent inflation of the non-rotating model during the MS takes place above an optical depth of $\tau \approx 1000$, being mainly driven by the sub-surface iron opacity peak. Importantly, this implies that predictions of observables such as effective temperature and spectral features during this phase rely on the accuracy of the atmosphere physics assumed at the very surface of the structure model. The main uncertainty herein lies in distinguishing between inflation and wind launching, for which evolutionary codes must make an ad-hoc assumption since the wind is not included in the structure models. To gain a deeper understanding of this regime, we therefore calculate dedicated atmosphere models smoothly connected to the deeper stellar structure in this work. 

Finally, we note that the MS inflation taking place here is fundamentally different from the expansion that a star experiences at the end of the MS, which is driven from much deeper inside the star (the convective core boundary lies at $\tau\sim 10^{10}$). In this regard, \citet{Sabhahit2024} distinguish between the terms ``inflation'' and ``expansion'', the former of which is used to describe the surface phenomenon discussed here. For consistency, we follow the same distinction throughout this paper. 

\section{Atmosphere models}
\label{sec:atm-models}

\subsection{The atmosphere in stellar evolution models}

Stellar evolution codes were originally not designed to treat stars with extended expanding atmospheres, such as VMS.  Moreover, due to a lack of hydrostatic equilibrium and very complex radiative transfer, realistic models for expanding atmospheres are numerically expensive on their own. Consequently, the surface boundary conditions in evolutionary codes are usually kept rather simple with very important atmosphere physics missing from the models. In GENEC, the upper layers of the star (representing between 1--2\% of the stellar mass) are counted as the so-called ``envelope''. In this region, the stellar structure is integrated hydrostatically over the density (instead of the pressure) in order to improve numerical stability and avoid density inversions\footnote{This actually marks one of the GENEC specifics as such density inversions are common results in other hydrostatic structure models \citep[see, e.g.,][]{Ishii1999,Graefener2012,Sanyal2015}.}.

At the top of the envelope, a gray atmosphere is appended as a boundary condition. The atmosphere stratification is then computed over the optical depth with $T^4(\tau) = \frac{3}{4}T_\mathrm{eff}^4(\tau + \frac{2}{3})$, where $T_\mathrm{eff}$ is the temperature of the layer immediately below the atmosphere, which is fixed at $\tau=\frac{2}{3}$. 

Notably, no wind is included in the structure of the atmosphere. However, \citet{Langer1989} derived a method of estimating a correction on the effective temperature based on a wind attached to the upper boundary of the stellar model. This method assumes a $\beta$ velocity law with $\beta = 2$ for the wind (see also Sect.\,\ref{sec:powr-models}) and computes the correction on the photospheric radius $R_{2/3}$ due to the additional optical depth of the wind as follows:
\begin{equation}
    \label{eq:r23approx}
    R_\mathrm{2/3} = R + \frac{3\dot{M}\kappa}{8\pi v_\infty}\text{,}
\end{equation}
where $\kappa$ is the (density-independent) opacity of the wind, $\dot{M}$ is the mass-loss rate and $v_\infty$ is the terminal wind velocity. $R$ is the surface radius of the stellar structure model. The calculated $R_\mathrm{2/3}$ is then the radius at which the wind reaches an optical depth of $2/3$. From this, the corrected effective temperature is obtained via the Stefan-Boltzmann relation. The opacity $\kappa$ in Eq.\,\eqref{eq:r23approx} should ideally be the flux-weighted mean opacity. However, as this cannot be determined from the structure models, it is either simply approximated by the electron-scattering opacity $\sigma_\text{e}$ \citep[e.g.\ in optical depth estimations such as in][]{Langer1989,Szecsi2015,Aguilera-Dena2022} or -- as described in more detail in \citet{Schaller1992} -- by a force multiplier approach $\kappa = \sigma_\text{e} \left( 1 + \mathcal{M}\right)$ with $\mathcal{M}$ following from a modified CAK \citep*[after][]{Castor1975} solution, namely the famous ``cooking recipe'' by \citet{Kudritzki1989}. The latter has become the standard in GENEC for the wind-corrected temperatures reported in the WR stage since \citet{Schaller1992}. Despite its level of sophistication, this ``wind correction'' requires not only the assumption of $\beta$ but also the choice of force multiplier parameters, which in the case of the GENEC application were taken from the $50\,$kK model in \citet{Pauldrach1986}.

In \citet{Groh2014a}, these wind-corrected effective temperatures (for $\beta = 2$ and the other assumptions) were compared to $T_\text{eff}$ values resulting from CMFGEN calculations for an evolutionary track of a star with an initial mass of $60\,M_\odot$. Generally, the GENEC-inherent corrections were found to be too strong with the specific differences being more complex than a simple scaling factor. Still, the basic concepts of the described procedure could prove useful when estimating effective temperatures of evolution models with significant winds. In Sect.\,\ref{sec:windcorr-teff}, we generalize it to any $\beta$ as well as the case of optically thin winds ($R_{2/3}<R$), and we compare the resulting approximation to results from detailed atmosphere models.

\subsection{Structure computation in detailed atmosphere models}
\label{sec:powr-models}

In this work, we use the Potsdam Wolf-Rayet code \citep[PoWR,][]{Graefener+2002,Hamann2003,Sander2015} to compute detailed models of the stellar atmosphere along a computed evolutionary track. PoWR models a spherically expanding non-LTE atmosphere in 1D, performing a detailed radiative transfer in the comoving frame (CMF). 

The structure of a stellar atmosphere within our framework is defined by a single function, either the density $\rho (r)$ or velocity $v(r)$, as the two are connected by the continuity equation
\begin{equation}
\label{eq:continuity}
    \dot{M} = 4\pi r^2 v(r)\rho (r)\text{.}
\end{equation}

In principle, the velocity structure in a 1D stationary atmosphere could be obtained consistently from solving the hydrodynamic equation of motion \citep[e.g.,][]{Graefener2005,Sander2017,Sundqvist2019}. However, coupled with a detailed non-LTE and CMF calculation, this is numerically very costly. While we will explore this approach for selected snapshots in a future paper -- based on the insights gained in this work -- an important first step is to connect structure and atmosphere models in a well-described manner using standard atmosphere modeling. In contrast to dynamically-consistent model calculations, this approach could later also be scaled to a larger parameter range.

In the standard approach, the structure is split radially into two domains. In the lower domain, we assume the kinetic acceleration $v\,\partial v/\partial r$ to be small compared to gravity, allowing us to neglect this term and assume hydrostatic equilibrium instead \citep[see][for the implementation in PoWR]{Sander2015}. This domain is then smoothly connected to the wind domain, for which we assume the velocity to follow a so-called beta-law:
\begin{equation}
    v(r) = v_\infty \left( 1+\frac{R_\ast}{r+p}\right)^\beta\text{,}
\end{equation}
where $v_\infty$ and $\beta$ are free parameters and $p$ is determined by the connection settings between the two domains. Two options for this connection are available in PoWR: Either, the wind velocity profile connects with a smooth gradient to the hydrostatic regime, or the connection is forced at a specific fraction of the sound speed. For the terminal velocity $v_\infty$, we choose $2.65$ times the escape velocity at the photosphere as suggested by (modified) CAK results \citep{Kudritzki2000}. Motivated by hydrodynamically-consistent models of WNh atmospheres by \citet{Graefener2008}, we choose $\beta=1$ \citep[see also Fig.\,2 in][]{Hamann2008}.

Our framework for the atmosphere models implies that the effective temperature at $\tau = \frac{2}{3}$, denoted simply as $T_\mathrm{eff}$, is no longer a boundary condition like in the evolutionary code, but is instead calculated intrinsically given the converged atmosphere and wind structure. As a result, $T_\mathrm{eff}$ can diverge significantly between different modeling methods given the same underlying stellar parameters.

\subsection{Grid setup}

\begin{table*}[]
\caption{Variable input parameters of the PoWR models along the evolutionary tracks of the non-rotating $150\,M_\odot$ models.}
\label{tab:input-params}
\centering
\begin{tabular}{ccccccccccccc}
grid & $t$   & $M$     & $\log (L/L_\odot)$  & $R_\mathrm{shallow}$   & $R_\mathrm{deep}$ & $\log (\dot{M} $ /  & $v_\infty$ & H      & C         & N         & O         & Ne        \\
 point         & Myr  & / $M_\odot$  &       &  / $R_\odot$     &  / $R_\odot$     &    $M_\odot\,\mathrm{yr})$      & / $\mathrm{km}\,\mathrm{s}^{-1}$  &        & / $10^{-4}$      & / $10^{-4}$      & / $10^{-4}$      & / $10^{-3}$      \\
          \hline\hline
00        & 0.02 & 149.7 & 6.382 & 17.23 & 15.69 & -4.831   & 4795 & 0.7200 & 23.11 & 6.588 & 57.34 & 2.029 \\
01        & 0.34 & 144.7 & 6.383 & 19.01 & 16.90 & -4.769   & 4481 & 0.7200 & 23.11 & 6.588 & 57.34 & 2.029 \\
02        & 0.61 & 139.6 & 6.384 & 20.41 & 17.67 & -4.701   & 4243 & 0.7200 & 23.11 & 6.588 & 57.34 & 2.029 \\
03        & 0.85 & 134.6 & 6.386 & 22.16 & 18.49 & -4.630   & 3992 & 0.7198 & 13.17 & 18.49 & 57.22 & 2.028 \\
04        & 1.04 & 129.7 & 6.387 & 23.37 & 18.94 & -4.583   & 3810 & 0.7016 & 0.7560 & 60.49 & 25.54 & 1.894 \\
05        & 1.23 & 124.7 & 6.388 & 24.39 & 19.16 & -4.545   & 3652 & 0.6702 & 0.9214 & 74.19 & 9.662 & 1.867 \\
06        & 1.40 & 119.5 & 6.390 & 25.36 & 19.24 & -4.513   & 3499 & 0.6322 & 0.9863 & 79.19 & 3.873 & 1.858 \\
07        & 1.56 & 114.6 & 6.392 & 26.40 & 19.27 & -4.484   & 3354 & 0.5917 & 1.013 & 80.80 & 1.990 & 1.850 \\
08        & 1.70 & 109.7 & 6.393 & 27.50 & 19.21 & -4.458   & 3207 & 0.5481 & 1.028 & 81.31 & 1.386 & 1.841 \\
09        & 1.84 & 104.5 & 6.395 & 28.77 & 19.02 & -4.431   & 3053 & 0.4993 & 1.041 & 81.46 & 1.197 & 1.832 \\
10        & 1.97 & 99.7  & 6.397 & 30.03 & 18.74 & -4.409   & 2909 & 0.4513 & 1.055 & 81.50 & 1.142 & 1.823 \\
11        & 2.10 & 94.7  & 6.399 & 31.85 & 18.39 & -4.384   & 2741 & 0.3994 & 1.066 & 81.50 & 1.117 & 1.812 \\
12        & 2.22 & 89.4  & 6.401 & 33.91 & 17.84 & -4.360   & 2563 & 0.3417 & 1.086 & 81.50 & 1.096 & 1.801 \\
13        & 2.33 & 84.6  & 6.404 & 36.42 & 17.23 & -4.338   & 2395 & 0.2889 & 1.103 & 81.50 & 1.076 & 1.791 \\
14        & 2.43 & 79.6  & 6.407 & 39.43 & 16.31 & -4.310   & 2209 & 0.2315 & 1.127 & 81.49 & 1.052 & 1.780 \\
15        & 2.53 & 74.7  & 6.414 & 36.04 & 13.54 & -4.266   & 2216 & 0.1740 & 1.156 & 81.48 & 1.022 & 1.769 \\ \hline
\end{tabular}
\tablefoot{The abundances in the last five columns are given as mass fractions.}
\end{table*}

Along the non-rotating $150\,M_\odot$ evolutionary track calculated with GENEC, we select 16 snapshots of the stellar structure along the MS for which to compute detailed atmosphere models using PoWR. These points, marked `00' (ZAMS) to `15' in Fig.\,\ref{fig:genec-hrd}, are separated by a mass loss of about $5\,M_\odot$ between them, which results in roughly even intervals in the HRD. The non-rotating track was selected (vs.\ the rotating one) for its notable inflation along the MS, a phenomenon we want to investigate further using our detailed atmosphere models. In order to explore the connection between the structure and atmosphere models as well as probe the effect of including the iron opacity peak within the atmosphere boundaries, we compute two models at each point with different connection depths.

In PoWR, the depth of the atmosphere is usually set via the parameter \texttt{TAUMAX}, which represents the Rosseland mean optical depth, without the contribution of line opacities, of the deepest radial point. This, together with the inner radius \texttt{RSTAR}, defines the lower boundary of the PoWR stratification. While the continuum optical depth is not a quantity directly available in stellar structure models, we can estimate it using Kramer's opacity law and extract the corresponding radial coordinate \texttt{RSTAR} from the GENEC structure. First, we compute a series of models with a relatively shallow optical depth of \texttt{TAUMAX=20}, which represents the ``standard'' way of computing atmosphere models. It is similar to the approach used by \citet{Schaerer1996a,Schaerer1996b}, \citet{Groh2014a} and \citet{Martins2022}. The latter can also serve as a comparison point for our results.

Alternatively to a connection based on optical depth, we can also define the lower boundary by its radial velocity $v_\mathrm{min}$. This enables us to connect the atmosphere structure continuously to the interior structure models at a specific density via the continuity equation. For the deep models, we choose the boundary as the radius at the density $\rho = 10^{-7}\,\mathrm{g}\,\mathrm{cm}^{-3}$, which is beneath the entire super-Eddington hot iron bump region. In comparison to the shallow models, the continuum optical depth of this point is on the order of $\sim$$1000$.

The stellar parameters extracted from GENEC are shown in Table \ref{tab:input-params}. In addition to the abundances shown in this table for H, C, N, O and Ne, we include the elements S, Si, and Ar, whose surface abundances we assume to be constant during the evolution and we obtain from \citet{Asplund2009}\footnote{The exact mass fractions were retrieved from this table: \url{http://astro.uni-tuebingen.de/~rauch/TMAP/TMAP_solar_abundances.html}}. The elements from Sc to Ni (the ``iron-group'') are combined in PoWR into a single generic element ``G'' in a super-level approach \citep{Graefener+2002}. Finally, the abundance of He is not an explicit input of the model but is instead calculated as the missing fraction from all the included elements.

For gridpoints 11 to 15, representing the last $0.4\,$Myr of MS evolution, the shallow modeling procedure did not produce a converged output. This is most likely due to the fact that the lower boundary of these models is beyond the Eddington limit, violating the requirement of hydrostatic equilibrium in the inner part. Furthermore, non-convergence could be a sign that the underlying (inflated) structure computed in GENEC has become so unphysical that it yields unrealistic stellar parameters, which are then used as input for PoWR.

\begin{figure*}
    \centering
    \includegraphics{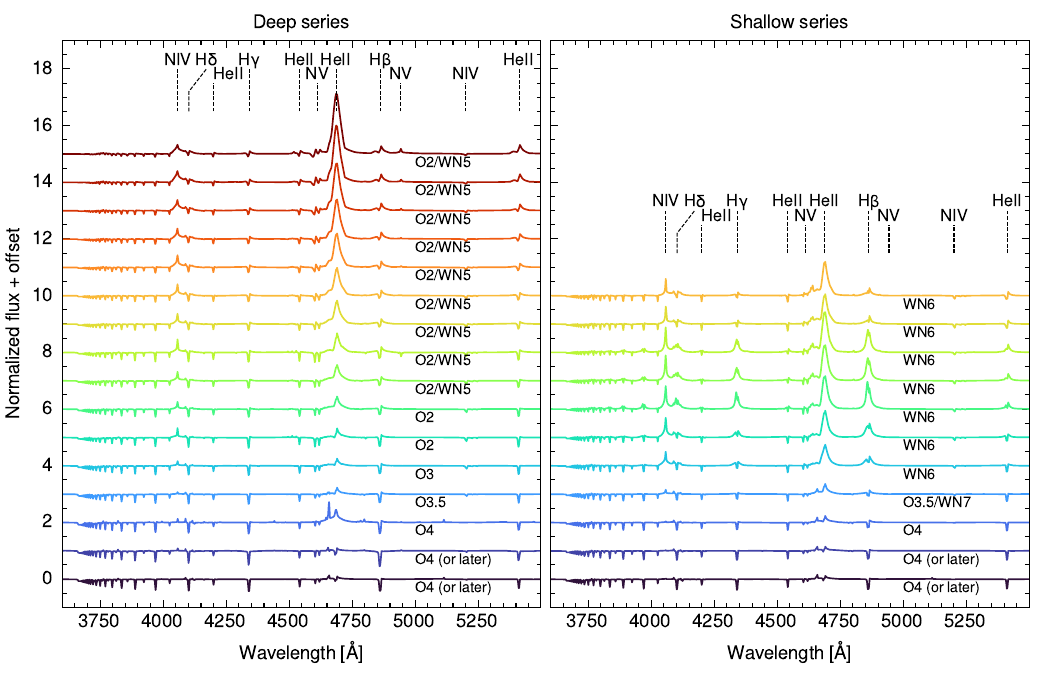}
    \caption{Optical spectrum for deep (left) vs. shallow (right) models across evolution (from bottom to top). The spectra are vertically offset so that the continuum level corresponds to the gridpoint number on the $y$-axis for easy reading. The spectral types were estimated based on the classification scheme in \citet{Crowther2011}.}
    \label{fig:opt-spectrum-evol}
\end{figure*}

\section{Spectroscopic analysis of the PoWR models}
\label{sec:shallow_models}

\subsection{Analysis of the evolutionary sequence of models}

Figure \ref{fig:opt-spectrum-evol} shows the evolution of the normalized optical spectra computed for the deep and shallow sequence  in PoWR. The ZAMS spectra (darkest blue) show very little difference between the two connection depths, with both models exhibiting mainly absorption features typical of O stars, as well as a weak \ion{He}{ii}\,4686 emission line. For the first four gridpoints, the spectra from both sequences look qualitatively very similar. 

After gridpoint 03 (between 0.85 and 1.04 Myr), there appears to be a switch in the spectral type of the shallow models, with several lines going into emission, notably H$\beta$ and H$\gamma$ among the other Balmer lines, as well as the \ion{N}{iv}\,4058 line, classifying the model as WNh-type. The emission lines become more prominent with time until reaching a peak at gridpoint 06, after which they begin to recede slightly, while still remaining stronger than those in the corresponding deep models. Such an evolution is consistent with spectra predicted using a similar method by \citet{Martins2022} for VMS. Interestingly, this behavior is completely absent in the deep model series, according to which we could classify this $150\,M_\odot$ star as O-type or at most O/WNh for its entire MS evolution, with the spectrum changing much more gradually.

While the hydrogen and nitrogen lines go into emission after the switch in the shallow models, the \ion{He}{ii} lines remain in absorption (with the exception of \ion{He}{ii}\,4686, whose emission increases considerably). Therefore, the spectral transition from O to WR type seems to be related only to the behavior of hydrogen in the atmosphere. To test this, we repeated the computation of the shallow gridpoint 04 spectrum while switching off the contribution of all Hydrogen lines. Indeed, this yields a qualitatively similar output to the ZAMS models. As it turns out, the appearance of hydrogen emission lines in the VMS is related to a switch in the ionization stratification of hydrogen in the star, which we discuss more in Sect.\,\ref{sec:structure}.

Interestingly, the transition from O to WR spectra in the shallow models series occurs much sooner than assumed in evolutionary codes, where a surface hydrogen mass fraction of $0.3$ is typically taken as the transition criterion. If the shallow models are to be believed, this suggests that a different criterion would be better suited for defining the beginning of the WR phase in evolution models for VMS.

Finally, we also inspected the spectrum in the UV range, shown in Fig.\,\ref{fig:uv-spectrum-evol}. Although there are some differences between the deep and shallow model series, they are much less striking than those discussed for the optical spectrum.

\subsection{Potential discrepancy in mass-loss diagnostics}

\begin{figure}
    \centering
    \includegraphics{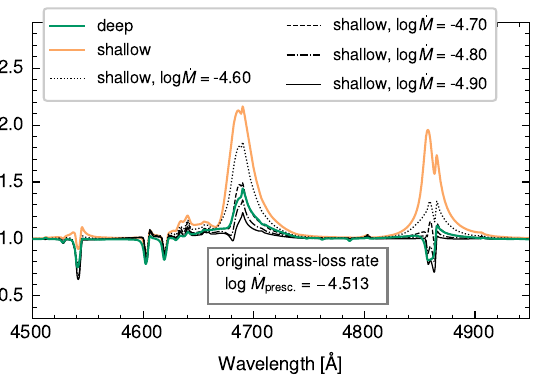}
    \caption{Fitting of the deep model spectrum with a shallow model of varying mass-loss rates at gridpoint 06. The original prescribed mass-loss rate used for the main models (green and yellow) and the evolutionary track is given in the boxed text for comparison.}
    \label{fig:mdot-fitting}
\end{figure}

Both the shallow and the deep model formally have the same mass-loss rates, despite their spectral differences. In turn, this raises the question whether the choice of the inner boundary in a model atmosphere could affect the derived mass-loss rates in quantitative spectroscopy. To test this, we now assume the deep model represents an observation which we would routinely fit with a shallow model, as those represent the typical atmosphere calculation assumptions. 

We carry out this experiment for gridpoint 06, where we visually identify the greatest difference in the two spectra (see Fig.\,\ref{fig:opt-spectrum-evol}). For this point, we take the existing shallow model as a baseline and then vary only the mass-loss rate until we obtain a model, the spectrum of which is close to that of the deep model at the same gridpoint. For simplicity, we fix the other input parameters during this procedure. While this would not be the case in an actual quantitative spectroscopic analysis, our main aim is to determine whether there is a systematic difference in the derived $\dot{M}$.

Figure \ref{fig:mdot-fitting} shows the spectra from the shallow and deep model as well as several adjusted versions of the shallow model with different mass-loss rates. We focus on the wavelength range around the \ion{He}{ii}\,4686 and H$\beta$ emission lines, since these tend to be strongly affected by the wind. For the inspected gridpoint, we see that reducing the mass-loss rate from the prescribed $-4.513$ to $-4.7$ results in the helium line matching the deep model's spectrum very closely. This reduction in $\dot{M}$ also results in the weakening of the H$\beta$ emission line, but not yet reaching the stronger absorption seen for the deep model. The models with $\log\dot{M} = -4.8$ and $\log\dot{M} = -4.9$ produce a closer fit to the H$\beta$ line. 

This simple modeling experiment reveals a possible uncertainty in the mass-loss rate determination of up to $0.4$\,dex, caused just by the connection of atmosphere models to evolution models. Ultimately, this underscores the still uncertain physics of VMS as implemented into structure and atmosphere models. If the deep atmosphere models were representative of the nature of VMS, this would imply that current fitting methods routinely underestimate the mass-loss rates of such stars.

\section{Structure comparison}
\label{sec:structure}

\subsection{Density profile}

\begin{figure*}
    \centering
    \includegraphics{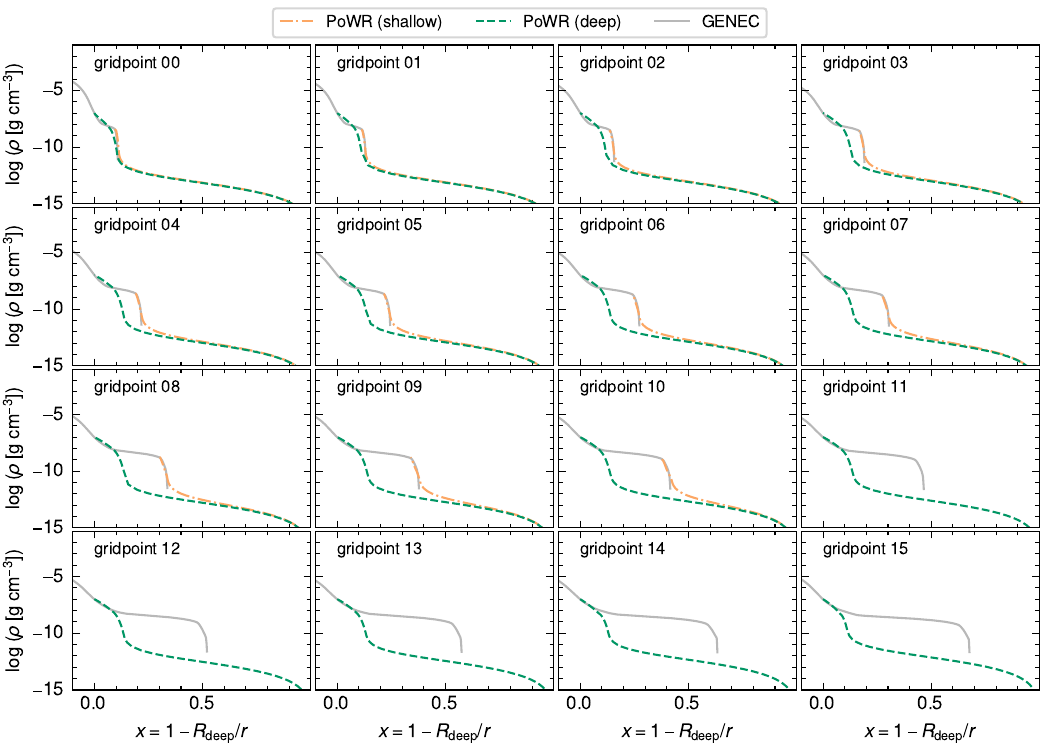}
    \caption{Density stratification across the MS evolution of the two PoWR model series compared to the GENEC structure models. The gridpoints correspond to the locations shown on the HRD in Fig.\,\ref{fig:genec-hrd}. The x-axis represents the modified radial coordinate $1-R_\mathrm{deep}/r$, where $R_\mathrm{deep}$ is the lower boundary radius $R_\ast$ of the deep model. This scaling allows us to visualize both the inner part and the wind domain of the atmosphere \citep[see, e.g., figures in][]{GonzalezTora2025, Debnath2024}.}
    \label{fig:powr-density-profile}
\end{figure*}

Figure \ref{fig:powr-density-profile} shows the density profiles of the GENEC model and both atmosphere models across the evolution of our example star. The evolution of the density profile of the GENEC models shows the establishment of a wide density plateau as a result of MS inflation (see also Fig.\,\ref{fig:interior_tau_evol}). By design, the deep PoWR model connects below this density plateau and the shallow model connects above it. As opposed to the GENEC models, the deep PoWR model does not include an inflated region. As a result, the density profiles of both atmosphere models differ increasingly over the evolution. However, although different in the inner part, the different density structures of the atmosphere models reconcile in the wind, past $x\approx 0.5$ ($r\approx 2R_\mathrm{deep}$). 

We also note that the density profiles of both PoWR models connect continuously to the GENEC model. For the deep models, this follows from tying their lower boundary directly to the GENEC density profile at $\rho = 10^{-7}\,\mathrm{g}\,\mathrm{cm}^{-3}$. For the shallow models, the density continuity indicates that our approximation of the continuum optical depth in GENEC is reasonable for the upper layers of the atmosphere. 

Given the differences between the obtained curves, it is a priory not clear which of the approaches marks the best representation of a VMS atmosphere. The limitation to 1D enforces approximations of the effect of the (hot) iron opacity peak, which differ between the structure and atmosphere model.

In GENEC, the model is assumed to be quasi-static instead of strictly static, meaning that the star is allowed to evolve on thermal timescales. As a result, hydrostatic equilibrium can still be found by allowing excess radiative energy to be converted to kinetic energy in the envelope, which ultimately causes the star to inflate. This inflation is a common effect in evolution models of stars close to the Eddington limit, where it also depends on the convection scheme applied in the sub-surface region \citep{Koehler2015,Graefener2021}. 

In contrast, the (quasi-)hydrostatic regime beneath the wind in stellar atmosphere codes does not allow any super-Eddington layer. The codes therefore either do not cover this regime and only model the wind or need to make an arbitrary cut of the radiative acceleration exceeding the Eddington limit. Following the tradition of TLUSTY \citep[e.g.,][]{Hubeny1995}, the default cut off in PoWR has long been at $\Gamma_\text{rad} = 0.9$, but recently increased to $0.99$. Any additional radiative acceleration is simply ignored, leading to a relatively compact atmosphere with a spectrum dominated by absorption lines. A priori, the evolution model may seem more physically plausible as it does not include an arbitrary cut-off of the Eddington parameter. However, this model also does not account for other effects that may play an important role related to the iron bump.

One possibility is to launch an optically thick wind already from the iron bump itself. If the star is close enough to the Eddington limit, acceleration over the iron bump can be sufficient to launch a supersonic flow also below the photosphere. This happens for example in a significant number of classical WR stars \citep[e.g.,][]{Graefener2005,Sander2020,Poniatowski2021}. In multiple dimensions, the redistribution of the gas into clumps with relatively high densities and channels with relatively low densities can artificially increase the effective driving in the atmosphere \citep{Moens2022}. This would mean that there is already a significant radial velocity field also underneath the photosphere which is not covered in a hydrostatic approach and as such strongly alter the atmosphere and wind physics. One direct effect of these altered physics is that the outflow will not follow the typically assumed $\beta$-velocity law. On the other hand, spectroscopic evidence suggests that launching a wind at the iron bump is not likely for VMS, given that wind-lines would then appear much stronger than observed \citep[see, e.g., the spectra in][arising from deep-wind launching]{Sander2023}. 

Alternatively, if the star does not manage to maintain an outflow, multi-dimensional models show that the atmosphere can fragment and partially fall down back on the star (sometimes known as a ``failed wind''). This suggests that the strong inflation seen in 1D models due to the effective scale height approaching infinity \citep[e.g.,][]{Langer2015} is actually unstable and an artifact of the current 1D model limitations combined with the forced hydrostatic equilibrium.
In the case of a failed wind, the iron bump can trigger radiatively-driven turbulent motion \citep{Moens2022, Debnath2024}. This turbulence is initiated by a gas parcel being driven upwards by increased opacity, expanding and therefore decreasing its opacity, causing it to fall back down. From current 3D model calculations, this effect is not expected to contribute much to the energy transport in the upper layers of hot stars, but could significantly affect its structure through the introduction of additional turbulent pressure. The turbulent pressure alters the total pressure scale height, which locally softens the temperature and density profiles. 
Some results from 2D hydrodynamical models indeed support the absence of a density plateau (or in more extreme cases an inversion), similar to our deep models \citep{Debnath2024}. In this case, our deep model sequence would represent a more realistic view of what VMS actually look like, which throws into question the interpretation of previous observations. 

Recent 1D approximations of O-star atmosphere calculations \citep{GonzalezTora2025} including a constant turbulent pressure show that even the density drop further inwards observed in our deep models might be washed out by turbulent motions. Therefore, it is likely that the density profiles of our two modeling approaches in PoWR represent two extreme possibilities for the atmospheric structure of VMS. Including a more sophisticated, i.e., non-radially constant, treatment of turbulent pressure in the 1D code should then result in a more gradual density decrease than our deep models, while not being as inflated as suggested by GENEC and the attached shallow model. Extending the implementation of turbulence and exploiting the code's full hydrodynamic capabilities is beyond the scope of the present paper, but will be  explored in a future work. 

\subsection{Temperature profile}

\begin{figure*}
    \centering
    \includegraphics{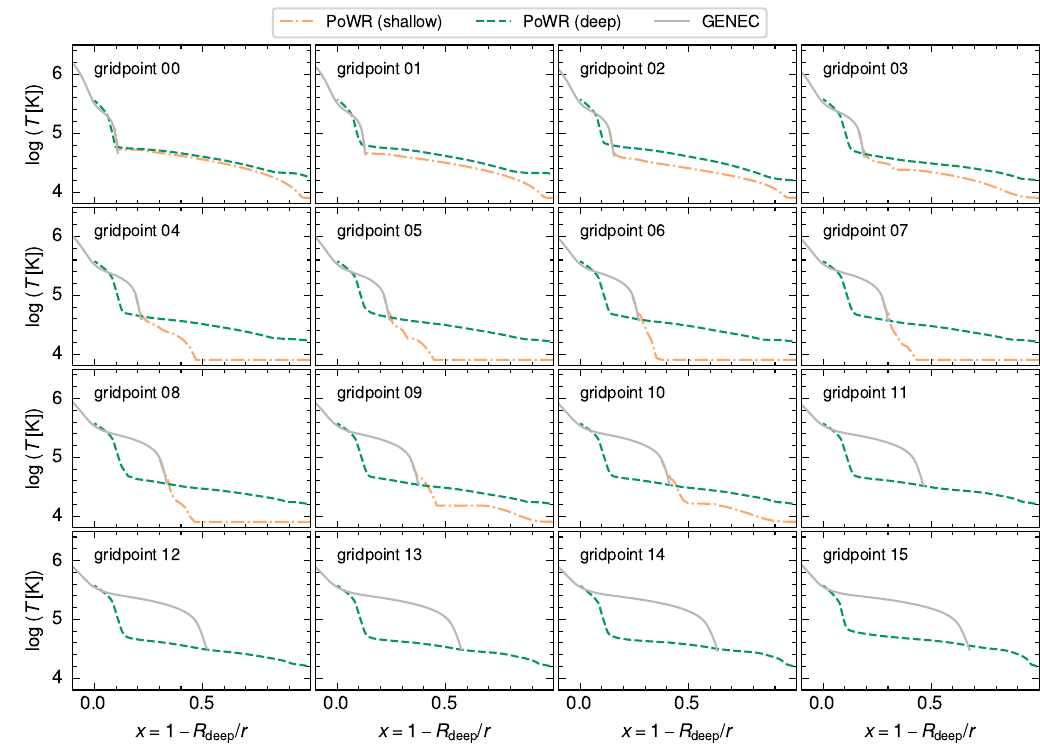}
    \caption{Temperature stratification across the MS evolution of the two PoWR model series as well as the GENEC structure models. The gridpoints correspond to the locations shown on the HRD in Fig.\,\ref{fig:genec-hrd}. For the PoWR models, the electron temperature is shown.}
    \label{fig:powr-temperature-profile}
\end{figure*}

\begin{figure}
    \centering
    \includegraphics{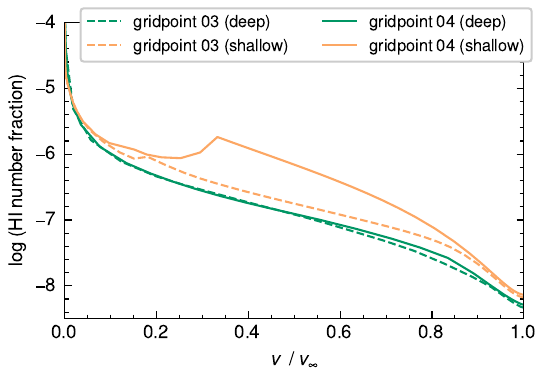}
    \caption{Stratification of the neutral hydrogen number fraction for the models at gridpoints 03 and 04 for the $150\,M_\odot$ track, showing the appearance of a recombination region for the shallow models at gridpoint 04. The subsequent shallow models in the series look qualitatively similar to the gridpoint 04 shallow model shown here, whereas no prominent recombination peak appears for any of the deep models.}
    \label{fig:hi-ion-transition}
\end{figure}

Figure \ref{fig:powr-temperature-profile} shows the evolution of the temperature profile of the GENEC model with the deep and shallow atmosphere model. For the PoWR models, the curves represent the electron temperature as there is no general temperature in a non-LTE regime. In the deep layers of the atmosphere, LTE is recovered and the electron temperature is thus identical to the gas temperature.

First, we note that the respective temperature connection between both atmosphere models and the structure model is excellent throughout the evolution. This is expected since the diffusion approximation is assumed to be valid at the lower boundary of the models, causing the electron temperature to coincide with the LTE temperature of the GENEC models at that point.

The qualitative temperature profile of the deep models remains very similar throughout the entire MS evolution. There is a sharp drop of the temperature until about $x\approx 0.1$ ($r\approx 1.1 R_\mathrm{deep}$) at $\sim$40--50 kK, after which the temperature decrease enters a more gradual regime. The transition point is optically very close to the surface of the star, at $\tau_\mathrm{Ross}=0.01$, and thus its spectrum is mostly formed in the region with the steep temperature gradient.

The shallow models follow a similar temperature profile as the deep models for the first four gridpoints. At gridpoint 04 the temperature decrease in the inner part suddenly starts to extend much deeper, reaching the minimum of 8000\,K imposed by PoWR after less than one stellar radius. This transition, coinciding with the spectral O$\rightarrow$WR transition (see Fig.\,\ref{fig:opt-spectrum-evol}), is closely related to the recombination of hydrogen in the wind, which we show in Fig.\,\ref{fig:hi-ion-transition}. As the evolution model inflates during the MS, the shallow PoWR models (which follow this inflation) are attached at an increasing radius and therefore have a decreasing lower boundary temperature. Eventually, this causes the recombination of \ion{H}{ii} to \ion{H}{i} in the atmosphere, which provides additional cooling leading to a further temperature decrease. This effect is a positive feedback loop, triggered by the crossing of some temperature threshold, which explains the very sudden change in temperature and ionization structure of the wind between just two gridpoints, representing an upper limit on the transition timespan of just 190\,kyr. 

Conversely, despite having the same mass-loss rate as shallow models, the temperature profile of the deep models keeps a very similar shape throughout the series with no further abrupt drops. For the same density, the temperature in the shallow models is much lower than in the deep ones. This could be a further sign that the inner boundary of the shallow models is not deep enough and some of the assumptions there are invalid. As a consequence, temperature correction algorithm likely overpredicts the cooling in the wind of the shallow models.

\subsection{Opacity profile}

\begin{figure*}
    \centering
    \includegraphics{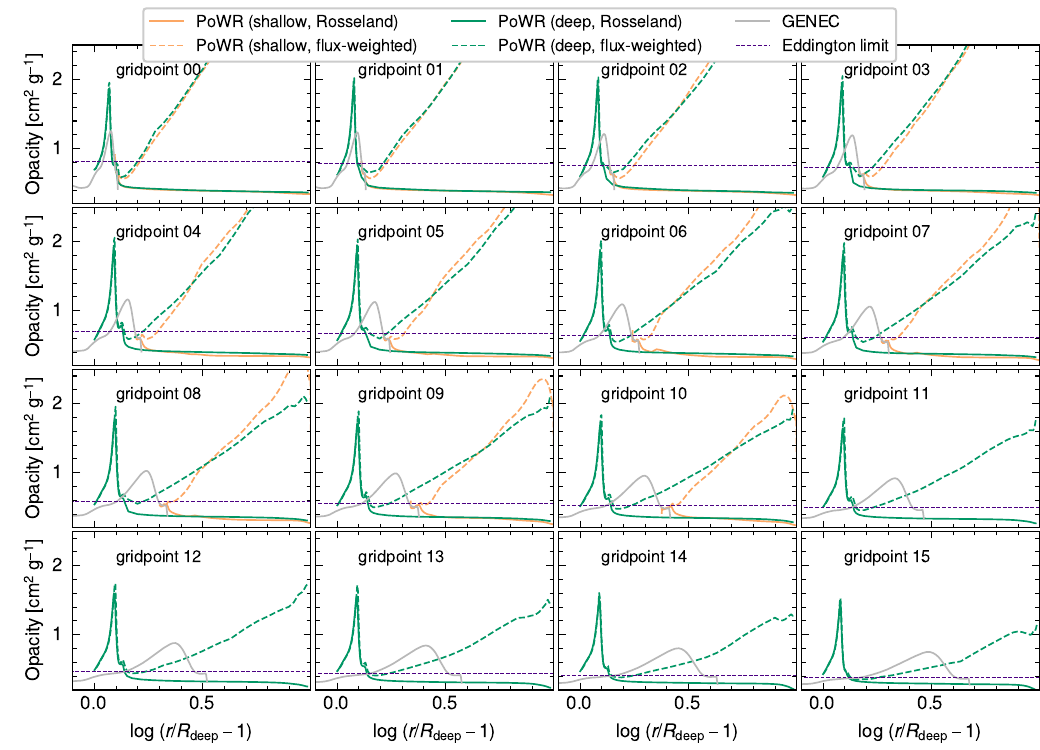}
    \caption{Evolution of the atmospheric profile of the Rosseland and flux-weighted mean opacities of the PoWR models, as well as the tabulated Rosseland mean opacity of the GENEC models. The opacity threshold for which the radiative acceleration surpasses gravity, as calculated by Eq.\,\ref{eq:kappa-edd}, is also shown.}
    \label{fig:kappa-profile}
\end{figure*}

Figure \ref{fig:kappa-profile} shows the opacity profile of the deep and shallow PoWR models and the GENEC structure models over the course of the MS evolution of our $150\,M_\odot$ VMS. For the GENEC models, the opacity is the Rosseland mean opacity obtained from tabulated values for a solar-mixture gas interpolated for metallicity, temperature and density. In PoWR, we obtain both the calculated Rosseland mean opacity given the conditions of the gas and the provided atomic data, as well as the flux-weighted mean opacity. As the name suggests, the latter is dependent on the spectral flux distribution and is directly related to the radiative acceleration via
\begin{equation}
    a_\mathrm{rad} = \frac{\kappa_\mathrm{F}L}{4\pi r^2c}\text{.}
\end{equation}
As such, we can also calculate the opacity corresponding to the Eddington limit from the condition $a_\mathrm{rad}=g$, which yields a constant only dependent on the current luminosity-to-mass ratio:
\begin{equation}
\label{eq:kappa-edd}
    \kappa_\mathrm{Edd} = 4\pi cG\frac{M}{L}
\end{equation}
The Rosseland mean opacity corresponds to the flux-weighted mean opacity in the case of a black-body spectrum (LTE), an approximation which is valid in the deeper layers of the star. In the wind, Fig.\,\ref{fig:kappa-profile} clearly shows a departure from LTE as the flux-weighted mean opacity becomes significantly super-Eddington, whereas the Rosseland mean opacity remains below the Eddington limit. In the deep PoWR models, this departure occurs above the super-Eddington iron bump region. 

In stellar structure models, which lack detailed radiative transfer needed to simulate a spectrum, the flux-weighted mean opacity (and thus the radiative acceleration) cannot be obtained intrinsically, rendering it impossible to self-consistently predict the launching of any line-driven stellar wind\footnote{A potential exception could potentially be winds launched by the (hot) iron bump, where line opacities not yet broadened by a wind velocity field are decisive. Moreover, flux-weighted opacity tables such as those from \citet{Poniatowski2022} might become a future option.}.

The profile of the Rosseland mean-opacity looks qualitatively similar throughout the MS evolution for both PoWR and GENEC models. Especially the deep PoWR model shows hardly any evolution in the opacity structure, with the location and shape of the hot iron bump remaining virtually unchanged. The iron bump itself has a noticeably higher maximum opacity in the PoWR model than predicted by the tabulated opacities in GENEC, in part due to the assumed Doppler broadening of lines in PoWR. For all models, the entire opacity profile shifts downwards slightly as the hydrogen mass fraction drops, reducing the contribution of Thompson scattering to the opacity. However, the Eddington limit decreases still faster due to strong mass loss and increasing luminosity, resulting in a stronger wind. For the later gridpoints, even the lower boundary of the deep PoWR models is already in the super-Eddington regime, in contrast to the GENEC models. In addition the opacity in GENEC is smeared out to larger radii as the atmosphere inflates, as also shown in Fig.\,\ref{fig:interior_tau_evol}.

In the wind regime, both PoWR models show a very similar opacity profile for the first grid points. For the later grid points, the flux-weighted opacity of the shallow models increases steeper in the wind than those of the deep models. This is a consequence of the recombination in the shallow models, which not only contributes hydrogen bound-free opacity, but also line opacity due to recombination of other elements, especially iron.

\subsection{Effective temperature in evolution vs.\ atmosphere models}

The effective temperature $T_\mathrm{eff}$ is a fundamental parameter of a massive star which plays a multi-faceted role in their modeling. In evolutionary codes, it is computed as a boundary condition necessary to solve the stellar structure equations and then provided as an output. Additionally, the value of $T_\mathrm{eff}$ feeds back into the evolution itself through its role in the computation of the mass-loss rate, which has important repercussions for many aspects of the evolution. Therefore, accurately modeling the effective temperature is not only important for the prediction of a star's location in the HRD, but also to constrain the nature of the evolution itself.

The effective temperature computed by GENEC is simultaneously the temperature at the surface of the structure model, which is a boundary condition specifically determined to obtain a valid solution of the stellar structure equations in the interior. By design, a gray atmosphere with an optical depth of exactly $2/3$ is appended to the structure model, resulting in the photosphere always coinciding with the surface of the hydrostatic star. However, this is not a physical necessity, and is in fact not true in stars with strong winds that completely obscure the hydrostatic layers.

In PoWR models, we refer to the effective temperature at $\tau=2/3$ as $T_{2/3}$. Note that this does not necessarily correspond to the electron or flux temperature at this layer, since the atmosphere is generally outside of LTE. Nonetheless, $T_{2/3}$ is physically most similar to the meaning of the effective temperature in evolutionary codes, which is why we choose this value as the point of comparison. We present the resulting effective temperatures from the different models in Table \ref{tab:teff-comparison}. The results are plotted in Fig.\,\ref{fig:t23-corrections}. It is evident that the deep models yield systematically higher effective temperatures, despite having the same mass-loss rates. This difference grows as the $T_\mathrm{eff}$ is predicted to decrease by GENEC (because of inflation), whereas it stays comparatively constant in the deep PoWR models.

\begin{table}[h!]
    \centering
    \caption{Output of effective photospheric temperature of the different models in kK.}
    \begin{tabular}{cccc}
    gridpoint & GENEC & PoWR (shallow) & PoWR (deep) \\
    \hline\hline
    00 &    54.4 & 54.3 & 54.5  \\
    01 &    51.8 & 51.8 & 52.0  \\
    02 &    49.9 & 49.9 & 50.9  \\
    03 &    47.9 & 47.8 & 49.4  \\
    04 &    46.6 & 46.5 & 48.9  \\
    05 &    45.6 & 45.5 & 48.2 \\
    06 &    44.7 & 44.5 & 48.4 \\
    07 &    43.8 & 43.6 & 48.2 \\
    08 &    42.8 & 42.6 & 48.1 \\
    09 &    41.8 & 41.5 & 48.6 \\
    10 &    40.8 & 40.4 & 49.0 \\
    11 &    39.5 & -- & 49.4 \\
    12 &    38.1 & -- & 49.9 \\
    13 &    36.6 & -- & 50.8 \\
    14 &    34.9 & -- & 51.6 \\
    15 &    36.3 & -- & 55.6 \\
        \hline
    \end{tabular}
    \label{tab:teff-comparison}
\end{table}

\subsection{A general correction method for the effective temperature in evolution models with wind}
\label{sec:windcorr-teff}

Starting from a computed structure model, we can cut off some upper layers and attach a simple wind regime to obtain a correction on the photospheric radius (and thus, temperature). Assuming a $\beta$ velocity law, this correction can be derived analytically based on simple approximations. We outline this method here. It constitutes a generalization on the wind correction derived by \citet{Langer1989} for optically thick winds with $\beta = 2$.

First we choose some radius $R_\mathrm{con}$ at which the wind connects to the hydrostatic regime. This choice is arbitrary and constitutes the only free parameter of this method. It can be chosen in a physically motivated way, and we will implement this method for different examples. The velocity law is then assumed to be
\begin{equation}
\label{eq:beta-law}
    v(r) = v_\infty\left(1-\frac{R_0}{r}\right)^\beta\text{,}
\end{equation}
where $R_0$ is a theoretical radius where the velocity law reaches $0$. Since continuity requires a strictly positive velocity, the connection point must be above the zero point, i.e. $R_\mathrm{con}>R_0$. Given a chosen $R_\mathrm{con}$, $R_0$ can be determined from the continuity equation as
\begin{equation}
    \frac{\dot{M}}{4\pi R_\mathrm{con}^2 \rho (R_\mathrm{con})} = v_\infty\left(1-\frac{R_0}{R_\mathrm{con}}\right)^\beta\text{,}
\end{equation}
where $\rho (R_\mathrm{con})$ is the density at the connection radius, which can be extracted from the stellar structure model. The wind parameters $\dot{M}$ and $v_\infty$ are the result of prescriptions.

The optical depth in the wind (i.e., for $r>R_\mathrm{con}$) as a function of radius can then be computed as an integral over opacity: 
\begin{equation}
    \tau (r) = \int_{r}^\infty \mathrm{\rho}(r')\kappa(r')\,\mathrm{d}r'\text{.}
\end{equation}
Using the continuity equation (Eq.\,\ref{eq:continuity}) to replace the density, inserting the velocity law (Eq.\,\ref{eq:beta-law}), and assuming a constant (Rosseland-mean) opacity $\kappa$ in the wind yields
\begin{equation}
    \tau (r) =\frac{\dot{M}\kappa}{4\pi v_\infty}\int_r^\infty \frac{1}{r'^2}\left(1-\frac{R_0}{r'}\right)^{-\beta}\mathrm{d}r'\text{,}
\end{equation}
which can be evaluated analytically to obtain
\begin{equation}
\label{eq:tau-general}
    \tau (r) = \frac{\dot{M}\kappa}{4\pi v_\infty R_0(1-\beta)}\left(1-\left(1-\frac{R_0}{r}\right)^{1-\beta}\right)\text{,}
\end{equation}
or in the case $\beta = 1$:
\begin{equation}
\label{eq:tau-beta1}
    \tau (r) = -\frac{\dot{M}\kappa}{4\pi v_\infty R_0}\ln \left(1-\frac{R_0}{r}\right)\text{,}
\end{equation}
Now, we can determine if this simple wind model is optically thick with the condition 
\begin{equation}
\label{eq:opt-thick-wind-condition}
    \tau_\mathrm{wind}:=\tau (r=R_\mathrm{con})>2/3\text{.}
\end{equation}
If this condition holds, then the photospheric radius is inside the wind and can be determined by simply inverting Eqs. (\ref{eq:tau-general}) and (\ref{eq:tau-beta1}) for $\tau (r)$ and setting $\tau = 2/3$. This yields
\begin{equation}
    R_{2/3} = R_0\left(1-\left(1-\frac{8\pi v_\infty R_0(1-\beta)}{3\dot{M}\kappa}\right)^\frac{1}{1-\beta}\right)^{-1}\text{,}
\end{equation}
and for $\beta = 1$:
\begin{equation}
    R_{2/3} = R_0\left(1-\exp\left(-\frac{8\pi v_\infty R_0}{3\dot{M}\kappa}\right)\right)^{-1}\text{.}
\end{equation}
If Eq.\,\eqref{eq:opt-thick-wind-condition} does not hold, then the wind is optically thin. In order to find the photospheric radius, we must then integrate the structure model below the connection point as follows:
\begin{equation}
    \tau (r) = \tau_\mathrm{wind} + \int^{R_\mathrm{con}}_r \rho(r')\kappa (r')\,\mathrm{d}r'\text{.}
\end{equation}
The photospheric radius $R_{2/3}$ can then determined by inverting this equation numerically and setting $\tau = 2/3$.

Finally, the corrected effective photospheric temperature can then be calculated as
\begin{equation}
    T_{2/3}=\left(\frac{L}{4\pi \sigma R_{2/3}^2}\right)^\frac{1}{4}\text{.}
\end{equation}

\begin{figure}
    \centering
    \includegraphics{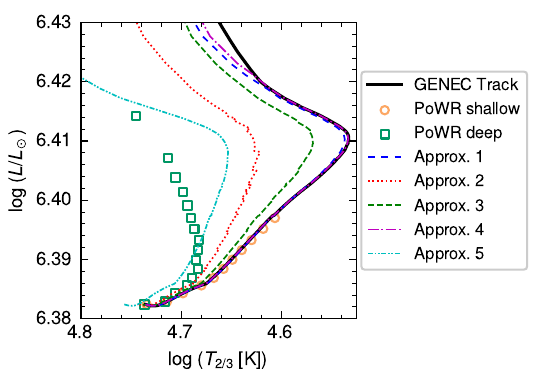}
    \caption{Evolution of the 150\,$M_\odot$ star on the HRD using different methods to obtain the effective photospheric temperature $T_{2/3}$. The solid black line shows the unmodified output of the GENEC track, and the square and circle scatter points show the output of the PoWR deep and shallow model series along the evolutionary track. The five non-solid lines follow the approximated wind-corrected photospheric temperature calculated from the GENEC structure models via the method outlined in Sect.\,\ref{sec:windcorr-teff} using different assumptions for the connection radius $R_\mathrm{con}$ (see main text for details).}
    \label{fig:t23-corrections}
\end{figure}

Using the described method, we can now determine a correction on the effective photospheric temperature computed by the evolutionary code and compare it with the output of the detailed atmosphere models computed by PoWR. Since the choice of the connection radius $R_\mathrm{con}$ remains a free parameter in the correction method, we show the results using different choices for this radius as examples:
\begin{itemize}
    \item[(1)] Connection where the velocity (computed via the continuity equation) reaches 20\% of the isothermal sound speed $v_\mathrm{s}=\sqrt{k_BT/m_H\mu}$.
    \item[(2)] Connection at the maximum of the Rosseland mean opacity in the hot iron bump.
    \item[(3)] Connection above the super-Eddington iron bump, where $\Gamma_\mathrm{Edd}$ falls below unity.
    \item[(4)] Connection at the surface radius determined by GENEC.
    \item[(5)] Connection at a Rosseland mean optical depth of 1000.
\end{itemize}

We show the resulting tracks alongside the output of the PoWR models in Fig.\,\ref{fig:t23-corrections}. It is clear that all correction methods yield a similarly shaped evolutionary track that is shifted towards the hot side of the HRD by differing amounts. Notably, none of the explored approximations can reproduce the track of the deep PoWR model series. While approximation (5) comes closest to these points, it still underestimates the effective temperature by up to 5\,kK in the second half of the MS. The mismatch between the GENEC approximations and the PoWR models is likely not only caused by the uncertainties of the approximation method itself, but also by the inconsistent treatment of the hydrostatic domain in PoWR causing differences between the two codes already below the wind. In the future, when hydrodynamically consistent series of PoWR models using more realistic physics exist as a point of comparison, this approximation method can be revisited to determine quick and reliable wind corrections from evolution models.

\section{Summary and conclusion}
\label{sec:summary}

In this work, we present two sequences of detailed atmosphere models computed in PoWR along an evolutionary track of a $150\,M_\odot$ VMS computed with GENEC. The two model sequences differ in their connection depth to the underlying structure models, allowing us to investigate the impact of this choice of boundary condition. The shallow models use a traditional approach, connecting at a continuum optical depth of $20$. For VMS evolution models, this is above a region experiencing significant opacity-induced inflation. Our deep atmosphere models use a new connection approach, employing a fixed density to connect to the structure models below this inflated region, representing the first such atmosphere models computed for VMS.

Our results show significant differences between the deep and shallow models in terms of spectral lines and atmospheric structure. Because PoWR suppresses super-Eddington opacities in the hydrostatic layers, no inflated region is established in the deep models, as most of the radiative force generated in the hot iron bump is ignored. Spectroscopically, this results in much more absorption-line dominated spectra typical of O-stars. Conversely, the shallow model spectra show a rapid transition from O-type to emission-dominated WNh-type spectra after around 1\,Myr of evolution, as previously expected of VMS. These differences in spectra between the two series arise despite all stellar parameters being identical, revealing previously unexplored systematic uncertainties in quantitative spectroscopy. For the mass-loss rate, we estimate a possible error of up to 0.4 dex (a factor of 2.5) arising from the choice of atmosphere model depth.

The deep and shallow models differ significantly in the deduced effective temperatures. The shallow models closely resemble the temperatures predicted by GENEC, while the deep models are up to $20\,$kK hotter. No simple correction formalism can fully account for this so far. In their current state, neither the evolutionary code nor our PoWR models (deep and shallow) are able to fully capture all of the relevant physics in the atmospheres of VMS. Insights from multi-dimensional atmosphere models have identified turbulent motion driven by the iron opacity peak as one such physical ingredient which has yet to be implemented consistently into 1D codes. We expect that a future introduction of turbulence in the atmosphere models and the usage of dynamically-consistent atmosphere models will result in a solution in-between our deep and shallow models.

\begin{acknowledgements}
   We would like to thank our referee whose comments helped us improve the quality of this manuscript. JJ acknowledges funding from the Deutsche Forschungsgemeinschaft (DFG, German Research Foundation) Project-ID 496854903 (SA4064/2-1, PI Sander). AACS, MBP, RRL, and VR are supported by the Deutsche Forschungsgemeinschaft (DFG - German Research Foundation) in the form of an Emmy Noether Research Group -- Project-ID 445674056 (SA4064/1-1, PI Sander).
   GGT acknowledges financial support by the Federal Ministry for Economic Affairs and Climate Action (BMWK) via the German Aerospace Center (Deutsches Zentrum f\"ur Luft- und Raumfahrt, DLR) grant 50 OR 2503 (PI: Sander).
   NM acknowledges the support of the European Research Council (ERC) Horizon Europe grant under grant agreement number 101044048.
   ECS acknowledges financial support by the Federal Ministry for Economic Affairs and Climate Action (BMWK) via the German Aerospace Center (Deutsches Zentrum f\"ur Luft- und Raumfahrt, DLR) grant 50 OR 2306 (PI: Ramachandran/Sander). JJ, MBP, RRL, and ECS are members of the International Max Planck Research School for Astronomy and Cosmic Physics at the University of Heidelberg (IMPRS-HD).
\end{acknowledgements}

\bibliographystyle{aa}
\bibliography{references}

\begin{thebibliography}{57}
\expandafter\ifx\csname natexlab\endcsname\relax\def\natexlab#1{#1}\fi

\bibitem[{{Aguilera-Dena} {et~al.}(2022){Aguilera-Dena}, {Langer}, {Antoniadis}, {Pauli}, {Dessart}, {Vigna-G{\'o}mez}, {Gr{\"a}fener}, \& {Yoon}}]{Aguilera-Dena2022}
{Aguilera-Dena}, D.~R., {Langer}, N., {Antoniadis}, J., {et~al.} 2022, \aap, 661, A60

\bibitem[{Asplund {et~al.}(2005)Asplund, Grevesse, \& Sauval}]{Asplund2005}
Asplund, M., Grevesse, N., \& Sauval, A.~J. 2005, in Astronomical Society of the Pacific Conference Series, Vol. 336, Cosmic Abundances as Records of Stellar Evolution and Nucleosynthesis, ed. I.~{Barnes}, Thomas~G. \& F.~N. {Bash}, 25

\bibitem[{{Asplund} {et~al.}(2009){Asplund}, {Grevesse}, {Sauval}, \& {Scott}}]{Asplund2009}
{Asplund}, M., {Grevesse}, N., {Sauval}, A.~J., \& {Scott}, P. 2009, \araa, 47, 481

\bibitem[{Bestenlehner {et~al.}(2014)Bestenlehner, Gr{\"a}fener, Vink, Najarro, de~Koter, Sana, Evans, Crowther, H{\'e}nault-Brunet, Herrero, Langer, Schneider, Sim{\'o}n-D{\'\i}az, Taylor, \& Walborn}]{Bestenlehner2014}
Bestenlehner, J.~M., Gr{\"a}fener, G., Vink, J.~S., {et~al.} 2014, \aap, 570, A38

\bibitem[{Castor {et~al.}(1975)Castor, Abbott, \& Klein}]{Castor1975}
Castor, J.~I., Abbott, D.~C., \& Klein, R.~I. 1975, \apj, 195, 157

\bibitem[{{Crowther} {et~al.}(2016){Crowther}, {Caballero-Nieves}, {Bostroem}, {Ma{\'\i}z Apell{\'a}niz}, {Schneider}, {Walborn}, {Angus}, {Brott}, {Bonanos}, {de Koter}, {de Mink}, {Evans}, {Gr{\"a}fener}, {Herrero}, {Howarth}, {Langer}, {Lennon}, {Puls}, {Sana}, \& {Vink}}]{Crowther2016}
{Crowther}, P.~A., {Caballero-Nieves}, S.~M., {Bostroem}, K.~A., {et~al.} 2016, \mnras, 458, 624

\bibitem[{{Crowther} \& {Walborn}(2011)}]{Crowther2011}
{Crowther}, P.~A. \& {Walborn}, N.~R. 2011, \mnras, 416, 1311

\bibitem[{{Cunha} {et~al.}(2006){Cunha}, {Hubeny}, \& {Lanz}}]{Cunha2006}
{Cunha}, K., {Hubeny}, I., \& {Lanz}, T. 2006, \apjl, 647, L143

\bibitem[{{Debnath} {et~al.}(2024){Debnath}, {Sundqvist}, {Moens}, {Van der Sijpt}, {Verhamme}, \& {Poniatowski}}]{Debnath2024}
{Debnath}, D., {Sundqvist}, J.~O., {Moens}, N., {et~al.} 2024, \aap, 684, A177

\bibitem[{Eggenberger {et~al.}(2008)Eggenberger, Meynet, Maeder, Hirschi, Charbonnel, Talon, \& Ekstr{\"o}m}]{Eggenberger2008}
Eggenberger, P., Meynet, G., Maeder, A., {et~al.} 2008, \apss, 316, 43

\bibitem[{Ekstr{\"o}m {et~al.}(2012)Ekstr{\"o}m, Georgy, Eggenberger, Meynet, Mowlavi, Wyttenbach, Granada, Decressin, Hirschi, Frischknecht, Charbonnel, \& Maeder}]{Ekstroem2012}
Ekstr{\"o}m, S., Georgy, C., Eggenberger, P., {et~al.} 2012, \aap, 537, A146

\bibitem[{{Gabrielli} {et~al.}(2024){Gabrielli}, {Lapi}, {Boco}, {Ugolini}, {Costa}, {Sgalletta}, {Shepherd}, {Di Carlo}, {Bressan}, {Limongi}, \& {Spera}}]{Gabrielli2024}
{Gabrielli}, F., {Lapi}, A., {Boco}, L., {et~al.} 2024, \mnras, 534, 151

\bibitem[{{Gonz{\'a}lez-Tor{\`a}} {et~al.}(2025){Gonz{\'a}lez-Tor{\`a}}, {Sander}, {Sundqvist}, {Debnath}, {Delbroek}, {Josiek}, {Lefever}, {Moens}, {Van der Sijpt}, \& {Verhamme}}]{GonzalezTora2025}
{Gonz{\'a}lez-Tor{\`a}}, G., {Sander}, A.~A.~C., {Sundqvist}, J.~O., {et~al.} 2025, \aap, 694, A269

\bibitem[{{Gormaz-Matamala} {et~al.}(2025){Gormaz-Matamala}, {Romagnolo}, \& {Belczynski}}]{GormazMatamala2025}
{Gormaz-Matamala}, A.~C., {Romagnolo}, A., \& {Belczynski}, K. 2025, \aap, 696, A72

\bibitem[{{Gr{\"a}fener}(2021)}]{Graefener2021}
{Gr{\"a}fener}, G. 2021, \aap, 647, A13

\bibitem[{{Gr{\"a}fener} \& {Hamann}(2005)}]{Graefener2005}
{Gr{\"a}fener}, G. \& {Hamann}, W.~R. 2005, \aap, 432, 633

\bibitem[{Gr{\"a}fener \& Hamann(2008)}]{Graefener2008}
Gr{\"a}fener, G. \& Hamann, W.~R. 2008, \aap, 482, 945

\bibitem[{{Gr{\"a}fener} {et~al.}(2002){Gr{\"a}fener}, {Koesterke}, \& {Hamann}}]{Graefener+2002}
{Gr{\"a}fener}, G., {Koesterke}, L., \& {Hamann}, W.~R. 2002, \aap, 387, 244

\bibitem[{{Gr{\"a}fener} {et~al.}(2012){Gr{\"a}fener}, {Owocki}, \& {Vink}}]{Graefener2012}
{Gr{\"a}fener}, G., {Owocki}, S.~P., \& {Vink}, J.~S. 2012, \aap, 538, A40

\bibitem[{Groh {et~al.}(2014)Groh, Meynet, Ekstr{\"o}m, \& Georgy}]{Groh2014a}
Groh, J.~H., Meynet, G., Ekstr{\"o}m, S., \& Georgy, C. 2014, \aap, 564, A30

\bibitem[{{Hamann} \& {Gr{\"a}fener}(2003)}]{Hamann2003}
{Hamann}, W.~R. \& {Gr{\"a}fener}, G. 2003, \aap, 410, 993

\bibitem[{{Hamann} {et~al.}(2008){Hamann}, {Gr{\"a}fener}, {Oskinova}, \& {Liermann}}]{Hamann2008}
{Hamann}, W.~R., {Gr{\"a}fener}, G., {Oskinova}, L., \& {Liermann}, A. 2008, in Astronomical Society of the Pacific Conference Series, Vol. 388, Mass Loss from Stars and the Evolution of Stellar Clusters, ed. A.~{de Koter}, L.~J. {Smith}, \& L.~B.~F.~M. {Waters}, 171

\bibitem[{{Higgins} {et~al.}(2023){Higgins}, {Vink}, {Hirschi}, {Laird}, \& {Sabhahit}}]{Higgins2023}
{Higgins}, E.~R., {Vink}, J.~S., {Hirschi}, R., {Laird}, A.~M., \& {Sabhahit}, G.~N. 2023, \mnras, 526, 534

\bibitem[{{Hubeny} \& {Lanz}(1995)}]{Hubeny1995}
{Hubeny}, I. \& {Lanz}, T. 1995, \apj, 439, 875

\bibitem[{{Ishii} {et~al.}(1999){Ishii}, {Ueno}, \& {Kato}}]{Ishii1999}
{Ishii}, M., {Ueno}, M., \& {Kato}, M. 1999, \pasj, 51, 417

\bibitem[{{Josiek} {et~al.}(2024){Josiek}, {Ekstr{\"o}m}, \& {Sander}}]{Josiek2024}
{Josiek}, J., {Ekstr{\"o}m}, S., \& {Sander}, A.~A.~C. 2024, \aap, 688, A71

\bibitem[{{K{\"o}hler} {et~al.}(2015){K{\"o}hler}, {Langer}, {de Koter}, {de Mink}, {Crowther}, {Evans}, {Gr{\"a}fener}, {Sana}, {Sanyal}, {Schneider}, \& {Vink}}]{Koehler2015}
{K{\"o}hler}, K., {Langer}, N., {de Koter}, A., {et~al.} 2015, \aap, 573, A71

\bibitem[{{Kudritzki} {et~al.}(1989){Kudritzki}, {Pauldrach}, {Puls}, \& {Abbott}}]{Kudritzki1989}
{Kudritzki}, R.~P., {Pauldrach}, A., {Puls}, J., \& {Abbott}, D.~C. 1989, \aap, 219, 205

\bibitem[{Kudritzki \& Puls(2000)}]{Kudritzki2000}
Kudritzki, R.-P. \& Puls, J. 2000, \araa, 38, 613

\bibitem[{{Langer}(1989)}]{Langer1989}
{Langer}, N. 1989, \aap, 210, 93

\bibitem[{{Langer} {et~al.}(2015){Langer}, {Sanyal}, {Grassitelli}, \& {Sz{\'e}sci}}]{Langer2015}
{Langer}, N., {Sanyal}, D., {Grassitelli}, L., \& {Sz{\'e}sci}, D. 2015, in Wolf-Rayet Stars, ed. W.-R. {Hamann}, A.~{Sander}, \& H.~{Todt}, 241--244

\bibitem[{{Martins} \& {Palacios}(2022)}]{Martins2022}
{Martins}, F. \& {Palacios}, A. 2022, \aap, 659, A163

\bibitem[{{Moens} {et~al.}(2022){Moens}, {Poniatowski}, {Hennicker}, {Sundqvist}, {El Mellah}, \& {Kee}}]{Moens2022}
{Moens}, N., {Poniatowski}, L.~G., {Hennicker}, L., {et~al.} 2022, \aap, 665, A42

\bibitem[{{Pauldrach} {et~al.}(1986){Pauldrach}, {Puls}, \& {Kudritzki}}]{Pauldrach1986}
{Pauldrach}, A., {Puls}, J., \& {Kudritzki}, R.~P. 1986, \aap, 164, 86

\bibitem[{{Poniatowski} {et~al.}(2022){Poniatowski}, {Kee}, {Sundqvist}, {Driessen}, {Moens}, {Owocki}, {Gayley}, {Decin}, {de Koter}, \& {Sana}}]{Poniatowski2022}
{Poniatowski}, L.~G., {Kee}, N.~D., {Sundqvist}, J.~O., {et~al.} 2022, \aap, 667, A113

\bibitem[{{Poniatowski} {et~al.}(2021){Poniatowski}, {Sundqvist}, {Kee}, {Owocki}, {Marchant}, {Decin}, {de Koter}, {Mahy}, \& {Sana}}]{Poniatowski2021}
{Poniatowski}, L.~G., {Sundqvist}, J.~O., {Kee}, N.~D., {et~al.} 2021, \aap, 647, A151

\bibitem[{{Ramachandran} {et~al.}(2018){Ramachandran}, {Hamann}, {Hainich}, {Oskinova}, {Shenar}, {Sander}, {Todt}, \& {Gallagher}}]{Ramachandran2018}
{Ramachandran}, V., {Hamann}, W.~R., {Hainich}, R., {et~al.} 2018, \aap, 615, A40

\bibitem[{Sabhahit \& Vink(2024)}]{Sabhahit2024}
Sabhahit, G.~N. \& Vink, J.~S. 2024, Stellar Expansion or Inflation?

\bibitem[{{Sabhahit} {et~al.}(2022){Sabhahit}, {Vink}, {Higgins}, \& {Sander}}]{Sabhahit2022}
{Sabhahit}, G.~N., {Vink}, J.~S., {Higgins}, E.~R., \& {Sander}, A. A.~C. 2022, \mnras, 514, 3736

\bibitem[{Salpeter(1955)}]{Salpeter1955}
Salpeter, E.~E. 1955, \apj, 121, 161

\bibitem[{{Sander} {et~al.}(2015){Sander}, {Shenar}, {Hainich}, {G{\'\i}menez-Garc{\'\i}a}, {Todt}, \& {Hamann}}]{Sander2015}
{Sander}, A., {Shenar}, T., {Hainich}, R., {et~al.} 2015, \aap, 577, A13

\bibitem[{{Sander} {et~al.}(2017){Sander}, {Hamann}, {Todt}, {Hainich}, \& {Shenar}}]{Sander2017}
{Sander}, A.~A.~C., {Hamann}, W.~R., {Todt}, H., {Hainich}, R., \& {Shenar}, T. 2017, \aap, 603, A86

\bibitem[{{Sander} {et~al.}(2023){Sander}, {Lefever}, {Poniatowski}, {Ramachandran}, {Sabhahit}, \& {Vink}}]{Sander2023}
{Sander}, A.~A.~C., {Lefever}, R.~R., {Poniatowski}, L.~G., {et~al.} 2023, \aap, 670, A83

\bibitem[{{Sander} {et~al.}(2020){Sander}, {Vink}, \& {Hamann}}]{Sander2020}
{Sander}, A. A.~C., {Vink}, J.~S., \& {Hamann}, W.~R. 2020, \mnras, 491, 4406

\bibitem[{{Sanyal} {et~al.}(2015){Sanyal}, {Grassitelli}, {Langer}, \& {Bestenlehner}}]{Sanyal2015}
{Sanyal}, D., {Grassitelli}, L., {Langer}, N., \& {Bestenlehner}, J.~M. 2015, \aap, 580, A20

\bibitem[{{Schaerer} {et~al.}(1996{\natexlab{a}}){Schaerer}, {de Koter}, {Schmutz}, \& {Maeder}}]{Schaerer1996a}
{Schaerer}, D., {de Koter}, A., {Schmutz}, W., \& {Maeder}, A. 1996{\natexlab{a}}, \aap, 310, 837

\bibitem[{{Schaerer} {et~al.}(1996{\natexlab{b}}){Schaerer}, {de Koter}, {Schmutz}, \& {Maeder}}]{Schaerer1996b}
{Schaerer}, D., {de Koter}, A., {Schmutz}, W., \& {Maeder}, A. 1996{\natexlab{b}}, \aap, 312, 475

\bibitem[{{Schaller} {et~al.}(1992){Schaller}, {Schaerer}, {Meynet}, \& {Maeder}}]{Schaller1992}
{Schaller}, G., {Schaerer}, D., {Meynet}, G., \& {Maeder}, A. 1992, \aaps, 96, 269

\bibitem[{{Schneider} {et~al.}(2018){Schneider}, {Sana}, {Evans}, {Bestenlehner}, {Castro}, {Fossati}, {Gr{\"a}fener}, {Langer}, {Ram{\'\i}rez-Agudelo}, {Sab{\'\i}n-Sanjuli{\'a}n}, {Sim{\'o}n-D{\'\i}az}, {Tramper}, {Crowther}, {de Koter}, {de Mink}, {Dufton}, {Garcia}, {Gieles}, {H{\'e}nault-Brunet}, {Herrero}, {Izzard}, {Kalari}, {Lennon}, {Ma{\'\i}z Apell{\'a}niz}, {Markova}, {Najarro}, {Podsiadlowski}, {Puls}, {Taylor}, {van Loon}, {Vink}, \& {Norman}}]{Schneider2018}
{Schneider}, F.~R.~N., {Sana}, H., {Evans}, C.~J., {et~al.} 2018, Science, 359, 69

\bibitem[{{Spera} \& {Mapelli}(2017)}]{Spera2017}
{Spera}, M. \& {Mapelli}, M. 2017, \mnras, 470, 4739

\bibitem[{Sundqvist {et~al.}(2019)Sundqvist, Bj{\"o}rklund, Puls, \& Najarro}]{Sundqvist2019}
Sundqvist, J.~O., Bj{\"o}rklund, R., Puls, J., \& Najarro, F. 2019, \aap, 632, A126

\bibitem[{{Sz{\'e}csi} {et~al.}(2015){Sz{\'e}csi}, {Langer}, {Yoon}, {Sanyal}, {de Mink}, {Evans}, \& {Dermine}}]{Szecsi2015}
{Sz{\'e}csi}, D., {Langer}, N., {Yoon}, S.-C., {et~al.} 2015, \aap, 581, A15

\bibitem[{{Upadhyaya} {et~al.}(2024){Upadhyaya}, {Marques-Chaves}, {Schaerer}, {Martins}, {P{\'e}rez-Fournon}, {Palacios}, \& {Stanway}}]{Upadhyaya2024}
{Upadhyaya}, A., {Marques-Chaves}, R., {Schaerer}, D., {et~al.} 2024, \aap, 686, A185

\bibitem[{{Vink}(2018)}]{Vink2018}
{Vink}, J.~S. 2018, \aap, 615, A119

\bibitem[{Vink {et~al.}(2001)Vink, de~Koter, \& Lamers}]{Vink2001}
Vink, J.~S., de~Koter, A., \& Lamers, H.~J.~G.~L.~M. 2001, \aap, 369, 574

\bibitem[{{Yusof} {et~al.}(2013){Yusof}, {Hirschi}, {Meynet}, {Crowther}, {Ekstr{\"o}m}, {Frischknecht}, {Georgy}, {Abu Kassim}, \& {Schnurr}}]{Yusof2013}
{Yusof}, N., {Hirschi}, R., {Meynet}, G., {et~al.} 2013, \mnras, 433, 1114

\bibitem[{{Zeidler} {et~al.}(2017){Zeidler}, {Nota}, {Grebel}, {Sabbi}, {Pasquali}, {Tosi}, \& {Christian}}]{Zeidler2017}
{Zeidler}, P., {Nota}, A., {Grebel}, E.~K., {et~al.} 2017, \aj, 153, 122

\end{thebibliography}

\appendix\onecolumn

\section{UV spectral evolution}
\begin{figure*}[h!]
    \centering
    \includegraphics{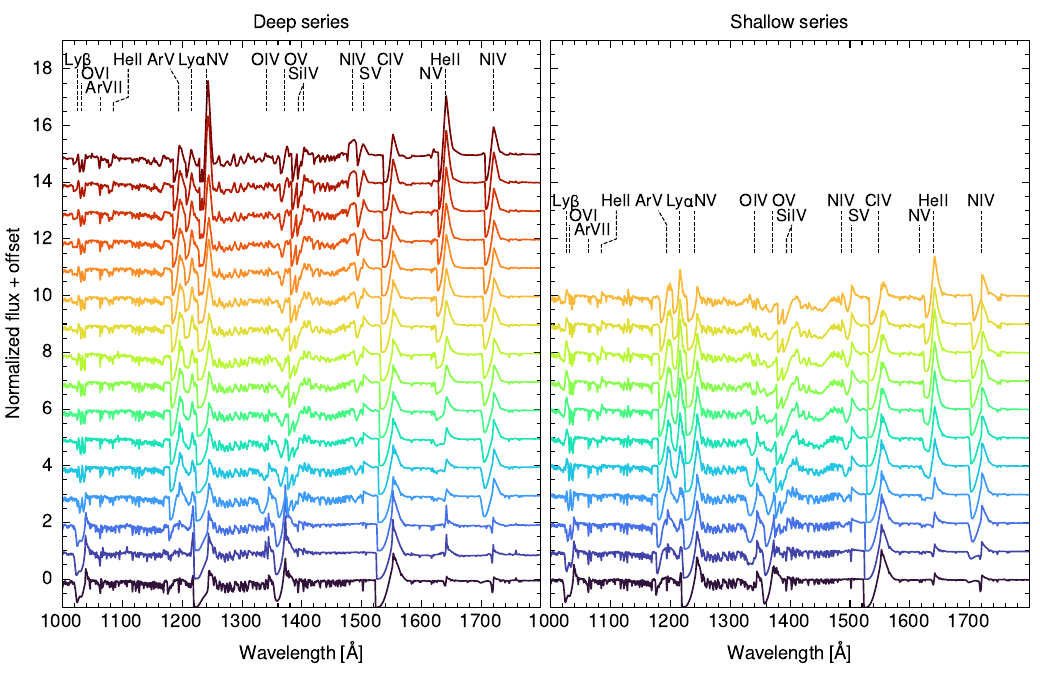}
    \caption{UV spectrum for deep (left) vs. shallow (right) models across evolution (from bottom to top).}
    \label{fig:uv-spectrum-evol}
\end{figure*}

\end{document}